\documentclass[pdftex,twocolumn,epjc3]{svjour3}     

\usepackage{epsfig,xfrac,booktabs,xfrac,amsmath,graphicx}
\usepackage{multicol}
\usepackage{multirow}

\RequirePackage[T1]{fontenc}

\smartqed  
\usepackage{amsmath}
\RequirePackage{graphicx}
\RequirePackage{mathptmx}     
\RequirePackage{flushend}
\RequirePackage[sort&compress,numbers]{natbib}
\RequirePackage[colorlinks,citecolor=blue,urlcolor=blue,linkcolor=blue]{hyperref}
 
\journalname{Eur. Phys. J. C}

\begin{document}

\title{Time evolution of Von Neumann entropy for a Kerr--Taub--NUT black hole} 

\author{Vicente A. Ar\'evalo\thanksref{e1,addr3}
        \and
        David Andrade\thanksref{e2,addr3}
        \and
        Clara Rojas \thanksref{e3,addr3}
}

\thankstext{e1}{e-mail: vicente.arevalo@yachaytech.edu.ec}
\thankstext{e2}{e-mail: dandrade@yachaytech.edu.ec}
\thankstext{e3}{e-mail: crojas@yachaytech.edu.ec}

\institute{Yachay Tech University, School of Physical Sciences and Nanotechnology, Hda. San Jos\'e S/N y Proyecto Yachay, 100119, Urcuqu\'i, Ecuador.\label{addr3}}

\date{Received: \today / Accepted: date}

\makeatletter
\renewcommand{\maketitle}{
  \twocolumn[\@maketitle]
  \@thanks 
}
\makeatother
\thispagestyle{empty}
\maketitle 

\begin{abstract}
In this work, we study the evolution of an evaporating black hole, described by the Kerr--Taub--NUT metric, which emits scalar particles. We found that allowing the black hole to radiate massless scalar particles increases the angular momentum loss rate while decreasing the loss rate of the NUT parameter and black hole mass. In fact, it means that angular momentum will disappear faster than the other black hole parameters (mass and NUT parameter) during the evaporation process. We also calculate the time evolution of the mass, angular momentum, and NUT parameter in order to get the evolution of the Von Neumann entropy of the black hole. We found that the entropy follows approximately the so-called Page curve, where the $\beta$ parameter, which quantifies the amount of radiation, affects the evaporation process. Implying that high $\beta$ values accelerate the evaporation process of a Kerr--Taub--NUT black hole.

\end{abstract}

\section{Introduction}\label{int}

The Kerr--Taub--NUT (KTN) space time is a stationary and axisymmetric vacuum solution of Einstein field equations \cite{chakraborty2014strong}, and it is described by three parameters: mass $M$, angular momentum $J$, and NUT parameter $l$. The presence of the parameter $l$ makes the metric have cosmic strings that are considered the source of the NUT parameter. According to Bonnor \cite{bonnor1969new}, the cosmic strings are physically regarded as a source of angular momentum. Therefore, the NUT parameter (NUT charge) quantifies the amount of angular momentum of the cosmic strings. When $l$ tends to zero, the KTN metric reduces to the Kerr space-time. 

The KTN metric can describe a black hole, and in the last few years, it has aroused much interest in the scientific community. For example, it has been theoretically examined the existence of the NUT parameter in M87$^{*}$ and Sgr A$^{*}$ by numerically deducing the shadow sizes in Kerr--Taub--NUT (KTN) space--time based on the data provided by the Event Horizon Telescope image \cite{ghasemi2021investigating,younsi2018electromagnetic,haroon2020shadow}. In the same way, the KTN space-time was used to explain the jet power and the radiative efficiency of black holes in different $X-ray$ binary systems \cite{narzilloev2023kerr}. Moreover, the Penrose process in KTN space-time was analyzed by Abdujabbarov et.al 2008 \cite{abdujabbarov2011penrose}. Morozova et.al 2008 \cite{morozova2008general} studied the magnetospheric structure surrounding a rotating, magnetized neutron star with a nonvanishing NUT (Newman--Unti--Tamburino) parameter.

It is a well-known fact that in 1975 Stephen Hawking theoretically found that black holes (BHs) are not totally black \cite{hawking1975particle}. Black holes can emit particles as Hawking radiation as a consequence of quantum mechanical effects near the event horizon of BHs. The emission of Hawking radiation leads to the evaporation of BHs and the information loss paradox since, at the end of the evaporation, it is not possible to have information about the matter swollen by the black hole. One possible solution to this paradox was proposed by Page in 1993 \cite{page1993information}. He considered that the information should be conserved if the black hole and its radiation are considered as a single pure quantum state that evolves under unitarity. Practically, he plotted the time evolution of the Von Neumman entropy for the quantum system in order to measure our incomprehension about the system, the so-called Page curve \cite{calmet2015quantum,almheiri2020page}.

Based on the previous facts, in the present paper we discuss the evaporation for a KTN black hole in order to give a description of its Page curve (i.e., time dependence of the Von Neumman entropy). The paper is organized as follows: In Section \ref{se1}, we describe the aspects of KTN space-time, including temperature and entropy. In Section \ref{se2}, we analyze the dynamics of a massless scalar field that is surrounding the KTN black hole. In Section \ref{se3}, we study the Hawking radiation, and in Section \ref{se4}, the time evolution of the black hole parameters and Von Neumman entropy are discussed. Finally, in Section \ref{se5}, we discuss the main conclusions. Throughout the work, we use solar masses as the unit of mass $M$, and to simplify the notation, we will use units such that $\hbar=c=G=k_{B}=1$,  being $k_{B}$ is Boltzmann's constant.

\section{Kerr--Taub--NUT Space-Time Structure}
\label{se1}

The metric that describes the KTN space-time is commonly expressed by the following line element \cite{Yang2020,Abdujabbarov2008,PhysRevD.108.103013,mukhopadhyay2004scalar,abdujabbarov2011penrose,miller1973global},

\begin{align}
\label{metric}
\mathrm{d} s^2= & -\frac{1}{\Sigma}\left(\Delta-a^2 \sin ^2 \theta\right) \mathrm{d} t^2\\
\nonumber
&+\frac{2}{\Sigma}\left[\Delta \chi-a(\Sigma+a \chi) \sin ^2 \theta\right] \mathrm{d} t \mathrm{d} \varphi \\
\nonumber
& +\frac{1}{\Sigma}\left[(\Sigma+a \chi)^2 \sin ^2 \theta-\chi^2 \Delta\right] \mathrm{d} \varphi^2+\frac{\Sigma}{\Delta} \mathrm{d} r^2+\Sigma \mathrm{d} \theta^2,
\end{align}
where parameters $\Sigma, \Delta$ and $\chi$ are given by the following expressions:

\begin{align}
\label{delta}
\Delta(r)&=r^2-2 M r-l^2+a^2,\\
\label{sigma}
\Sigma(r,\theta)&=r^2+(l+a \cos \theta)^2,\\
\chi(\theta)&=a \sin ^2 \theta-2 l \cos \theta,
\end{align}
and $M$, $a$, $l$ are the mass, rotation parameter, and NUT parameter, respectively, with $r\in [0,+\infty]$, $t\in [0,+\infty]$, $\theta \in [0, \pi]$, and $\varphi \in [0,2\pi]$.

It is clear that when $l=0$, the metric \eqref{metric} represents a Kerr space--time. Additionally, when $a=0$, it reduces to Schwarzschild space-time.

The KTN space-time describes a black hole, and the radial coordinate of the event horizon can be obtained by setting equation \eqref{delta} to zero. Therefore, it is possible to have the following result:

\begin{equation}
\label{horizons}
r_{\pm}= M \pm \sqrt{M^2+l^2-a^2}.
\end{equation}

Now, from Eq. \eqref{horizons}, it is possible to have the event horizon coordinate.

\begin{equation}
\label{horizon}
r_{h}=r_{+}= M + \sqrt{M^2+l^2-a^2}.
\end{equation}

In the same way, using the temporal part of Eq. \eqref{metric} and equal it to zero, we have the static limit, given by

\begin{equation}
\label{ergo}
r_{sta}= M + \sqrt{M^2+l^2-a^2\cos ^2 \theta},
\end{equation}
which practically defines the ergoregion, the place where the objects tend to move in the direction of the black hole's rotation. 

The region between the static limits Eqs. \eqref{horizon} and \eqref{ergo} is known as the ergoregion. Within it, it is not possible to ``stand still''. \cite{carroll2019spacetime,ferrari2020general, podolsky2020accelerating}.

Additionally, in terms of dimensionless parameters, setting $a_{\ast}=\frac{a}{M}$ and  $l_{\ast}=\frac{l}{M}$, the equations \eqref{horizon} and \eqref{ergo} are given as follows:

\begin{align}
r_{h}&= M\left(1 + \sqrt{1+l_{\ast}^2-a_{\ast}^2}\right),\label{horizon1}\\
r_{erg}&= M \left(1+ \sqrt{1+l_{\ast}^2-a_{\ast}^2\cos ^2 \theta}\right),\label{ergo1}
\end{align}

The quantities described in expressions \eqref{horizon1} and \eqref{ergo1} are helpful for astrophysical purposes \cite{Chakraborty2019}. Moreover, it is important to mention that to obtain the essential singularity, we need to consider that $g_{tt}\rightarrow \infty$, which implies that  $\Sigma=0$  which is given by the expression \eqref{sigma}. Then, it is possible to have the following results \cite{Yang2020,Chakraborty2019,PhysRevD.108.103013}:

\begin{equation}
\label{singular}
r=0\quad  \text{and} \quad \theta=\cos ^{-1}\left( -\dfrac{l_{\ast}}{a_{\ast}}\right).
\end{equation}

Now, it is important to mention that to have a black hole, the following condition, $a^2\geq l^2$ or $a_{\ast}\geq l_{*}$ \cite{Chakraborty2019,mukherjee2019some,Yang2020}, must be accomplished: Otherwise, we will have a regular black hole. Moreover, the above condition allows us to simplify the metric as a flat space at infinity.

\subsection{Temperature}

As a consequence of the Hawking radiation, black holes must have a non--zero temperature. In this case, we calculate the temperature of a KTN black hole. 

Generally, the temperature of a black hole is given by the following expression:

\begin{equation}
\label{temperature}
T_{H}=\frac{\kappa}{2 \pi},
\end{equation}
where $\kappa $ is known as the surface gravity  (see \cite{ferrari2020general} for more details).

In order to calculate the temperature for a KTN black hole, we need to use Eq. 
\eqref{temperature}, and the surface gravity calculation provided by Siahaan \cite{siahaan2021magnetized}.

\begin{equation}
\label{gra}
\kappa= \frac{r_{+}-r_{-}}{2\left( r_{+}^2+a^2+l^2\right)},
\end{equation}
where $r_+$ and $r_-$  (expressions \eqref{horizons}) are the outer and inner horizons of the black hole, respectively. So that, the temperature is described as follows:

\begin{equation}
\label{KNT temp}
T_{KTN}=\dfrac{r_{+}-r_{-}}{4 \pi \left( r_{+}^2+a^2+l^2\right)}.
\end{equation}

If we replace the expressions \eqref{horizons} in Eq. \eqref{KNT temp}, we can get the temperature of the KNT black hole, which is expressed as follows \footnote[1]{Note that this expression reduces to the temperature of a Schwarzschild black hole when $a\rightarrow 0$ and $l\rightarrow 0$.}:

\begin{equation}
\label{KNT main}
T_{KTN}= \dfrac{\sqrt{M^2+l^2-a^2}}{2 \pi  \left[\left(\sqrt{M^2+l^2-a^2}+M\right)^2+a^2+l^2\right]}.
\end{equation}

\subsection{Black Hole Entropy}
\label{sarea}

If black holes have temperature, they must have entropy, which is given by the following expression

\begin{equation}
\label{bekenstein}
S_{BH}=\dfrac{A}{4},
\end{equation}
where $A$ is the area of the black hole event horizon which can be calculated as follows:

\begin{equation}
\label{integral}
A_{H}=\int_{r=r_{h}} \sqrt{g_{\theta \theta}g_{\varphi \varphi}}\mathrm{d}\theta \mathrm{d}\varphi.
\end{equation}

In order to quantify the knowledge about the Kerr--Taub--NUT black hole, it is necessary to calculate the entropy associated with the black hole.
For this purpose, we use the formula \eqref{bekenstein}, Eq. \eqref{integral}, and the corresponding elements of the metric \eqref{metric}; considering that at the event horizon $\Delta(r_{+})=0$. Thus,

\begin{equation}
\label{area}
A_{KTN}=\int_{0}^{2\pi}\int_{0}^{\pi} (\Sigma+a \chi) \sin  \theta \mathrm{d}\theta \mathrm{d}\varphi= 4\pi(\Sigma+a \chi).
\end{equation}

Therefore, taking the previous result (Eq. \eqref{area}), the entropy is given by the following expression:

\begin{equation}
\label{entropy2}
S_{KTN}=2\pi \left(M^2+l^2+M \sqrt{M^2+l^2-a^2}\right).
\end{equation}

\section{Massless Scalar Field in Kerr--Taub--NUT Space--Time}\label{se2}

The dynamics of a minimally coupled massless scalar field $\phi$ in the Kerr--Taub--NUT spacetime is described in terms of the Klein--Gordon equation for a curved spacetime

\begin{equation}
\label{gordon1}
g^{\mu \nu}\nabla _{\mu}\partial_{\nu}\phi=0,
\end{equation}
where $g^{\mu \nu}$ is the metric that describes the spacetime.

The most common method to solve Eq. \eqref{gordon1} is applying the separation of variables method, which makes the problem more tractable \cite{PhysRevD.97.084044,PhysRevLett.29.1114,1973ApJ...185..635T}.

Therefore, the scalar field must be decomposed as follows:

\begin{equation}
\label{decomposition}
\phi(t, r, \theta, \varphi)=e^{-i \omega t} R_{\ell m}(r) S_{\ell m}(\theta) e^{i m \varphi},
\end{equation}
where $\omega$, $\ell$, and $m$ denote the frequency of the $j-$th emission particle, the spherical harmonic quantum number, and the axial quantum number, respectively \cite{don13particle}. 

Once the decomposition \eqref{decomposition} is applied to Eq. \eqref{gordon1}, we can obtain the radial part of Klein--Gordon equation, which is commonly expressed as the Teukolsky equation, which is useful to describe Kerr black hole perturbations\cite{1973ApJ...185..635T,page1976particle, hartle1974analytic}. The Teukolsky equation is described as follows:

\begin{align}
\label{teukolsky}
&\frac{1}{\Delta^s} \dfrac{\mathrm{d}}{\mathrm{d} r}\left(\Delta^{s+1} \dfrac{\mathrm{d} R_{\ell m}}{\mathrm{d} r}\right)\\
\nonumber&+\left[\dfrac{K^2-2 i s(r-M) K}{\Delta}+4 i s \omega r-\lambda_{\ell m}\right] R_{\ell m}=0,
\end{align}
where $\Delta$ is a function that comes from the metric, $K$ is an expression that depends on the metric, and $s$ is the spin weight of the field and takes the values $0$, $1$ or $2$ for scalar, electromagnetic, or gravitational perturbations \cite{seidel1989comment}. 

As we will work with scalar perturbations, it is essential to use a spin weight $s=0$. Moreover, considering $K$ used by Yang \textit{et al.} \cite{yang},  Eq. \eqref{teukolsky}, which represents the radial part of the perturbations for a KTN black hole, will be given by the following equation:

\begin{equation}
\label{radial}
\dfrac{\mathrm{d}}{\mathrm{d}r}\left( \Delta \frac{\mathrm{d} R_{\ell m}}{\mathrm{d}r}\right)+\left( \frac{K^{2}}{\Delta}-\lambda_{\ell m}\right)R_{\ell m}=0,
\end{equation}
where $\lambda_{\ell m}$ is the separation constant and is given by $\ell(\ell+1)+\mathcal{O}(a^2 \omega^2)$ 
\cite{seidel1989comment}, $\Delta$ is given by \eqref{delta} and 

\begin{equation}
\label{g}
K=\omega\left(r^2+l^2+a^2\right)-a~ m.
\end{equation}

It is noteworthy to emphasize that leveraging the radial component of the Teukolsky equation enables a more efficient representation of the radial part of the Klein--Gordon equation \eqref{gordon1} governing the scalar field. Furthermore, we tailor this equation to our specific metric, capitalizing on the Kerr--like nature of KTN space-time. 

\subsection{Effective potential}
\label{efe}

Once we have the radial part of the  perturbation, let us use Eq. \eqref{radial} in order to calculate the effective potential.

First of all, it is important to obtain the Schr\"odinger-like wave equation. For this, let us  consider the following change of variable:

\begin{equation}
\label{variable}
R_{lm}=\dfrac{U}{\sqrt{r^2+a^2+l^2}},
\end{equation}
and the tortoise coordinate

\begin{equation}
\label{tortoise}
\dfrac{\mathrm{d}r}{\mathrm{d}r_{\ast}}=\dfrac{\Delta}{(r^2+a^2+l^2)}.
\end{equation}

After applying the above conditions to Eq. \eqref{radial}, we got the following Schr\"odinger--like wave equation for a KTN black hole

\begin{equation}
\label{wave1}
\dfrac{\mathrm{d}^2 U(r_{\ast})}{\mathrm{d} r_{\ast}^2}+\left({\omega^2}-V_{e f f}\right)U(r_{\ast})=0,
\end{equation}
where the effective potential for a KTN black hole in an massless scalar field $V_{eff}$ is expressed as follows;

\begin{align}
\label{potential}
\nonumber
V_{eff}=V_{\omega \ell m}(r)&=\dfrac{2 ~a ~m~ \omega}{a^2+l^2+r^2}-\dfrac{a^2 m^2- \ell (\ell+1)\Delta }{\left(a^2+l^2+r^2\right)^2}\\
&-\dfrac{\Delta  r-\Delta  \left(2 r^2-2 M r\right)-\Delta ^2}{\left(a^2+l^2+r^2\right)^3}-\frac{2 \Delta ^2 r^2}{\left(a^2+l^2+r^2\right)^4}.
\end{align}

In general, Eq. \eqref{potential} represents the potential that the massless scalar field perturbations (bosons) will feel in the space-time that is surrounding the black hole.

\begin{figure*}[th]
\centering
\includegraphics*[width=1\textwidth]{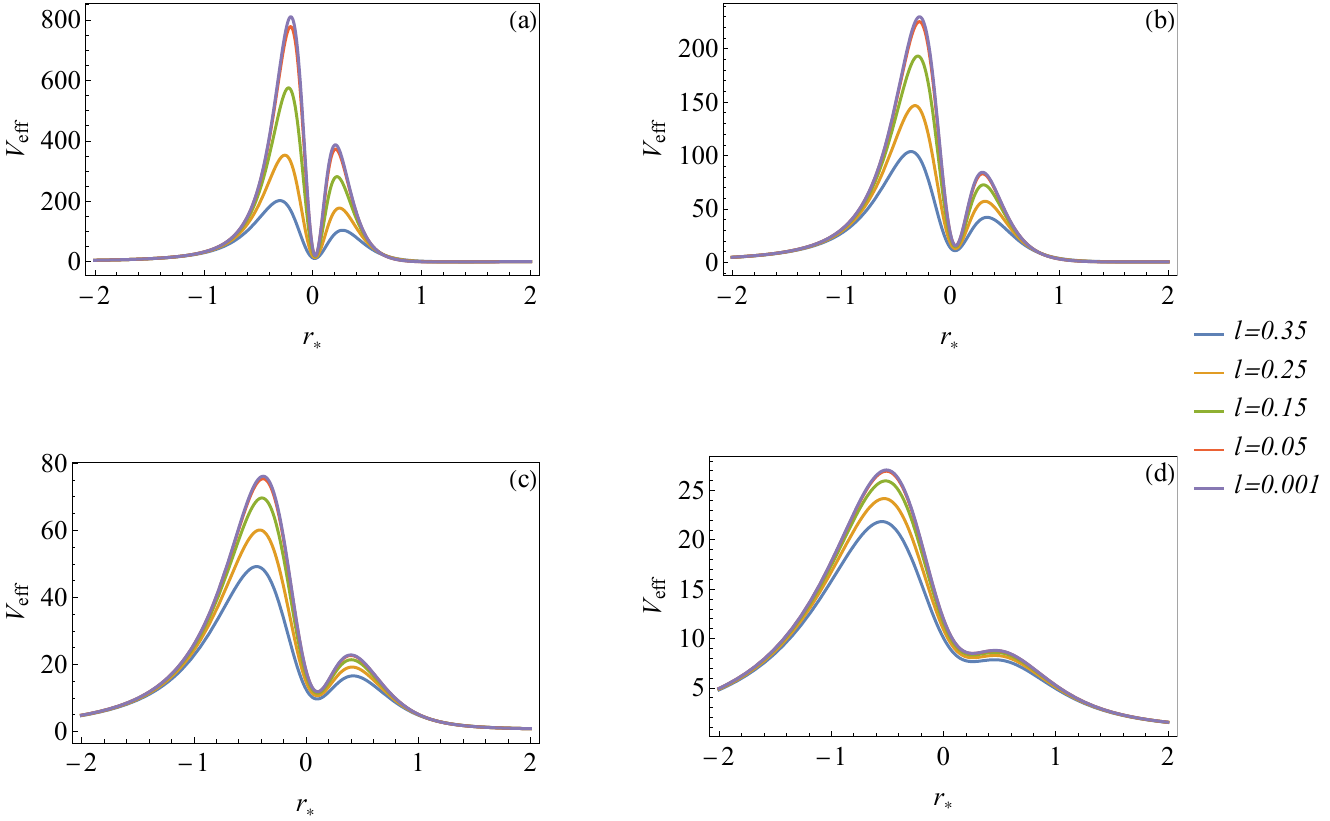}
\caption{Plots of $V_{eff}$ \eqref{potential} for various values of the NUT parameter $l$. Here the parameters are $M=3$,~  $\omega=5$, $\ell =m=1$, (a) $a= 0.35$, ~ (b) $a= 0.5$, ~ (c) $a= 0.7$, ~ and (d) $a= 1$.
}\label{Fig7}
\end{figure*}

In the figure \ref{Fig7}, we see different profiles for the effective potential  \eqref{potential} in terms of the tortoise coordinate $r_{\ast}$. First of all, it is important to say that the asymptotic behavior of the potential at $+\infty$ and $-\infty$ is zero, which guarantees that we can approximate the solutions of the Eq. \eqref{radial} as a combination of outgoing and ingoing waves to the black hole horizon \cite{panotopoulos2018greybody,leite2019axis}. But we are not interested in this fact because we are going to use the potential to calculate the GFs instead of the wave solutions. An important aspect of the potential is that it looks like a composition of Gaussian--like shapes, which is a characteristic of asymptotically flat spacetimes \cite{panotopoulos2018greybody}. 

Another relevant feature of the effective potential \eqref{potential} is that the NUT parameter is related to the height of the potential. For low values of the NUT parameters, it is clear that the intensity of the potential increases, and for high values of the NUT parameter, the potential decreases. Now, if we consider the value of the rotation parameter, it is noticeable that for slow rotations, the potential is intense, peaked, and narrow Fig \ref{potential}~(a). On the other hand, when the rotations increase, it is possible to see that the intensity and narrowness decrease; this aspect contributes to the smoothness of the potential profile Fig. \ref{potential}~(b), (c). Finally, if we consider a maximally rotating KTN black hole Fig. \ref{potential}~(d), it is possible to identify a smooth potential with low intensity compared to the other situations. Moreover, when the rotation increases, the peaks get broader. In general, it is possible to say that when both the rotation and the NUT parameter of the black hole increase, the potential gets less intense and smoother, allowing the particles to escape more easily from the event horizon than when the NUT parameter and the rotation parameter are small.


\section{Hawking Radiation for a KTN Black Hole}\label{se3}

Black holes emit radiation as greybodies as a consequence of the greybody factors (GFs) that modify the blackbody nature of the Hawking radiation spectrum. 
The greybody factors, or the absorption cross-sections, are  frequency-dependent factors that measure the modification of the original blackbody radiation \cite{Panotopoulos2018}. A high value of GFs indicates a high likelihood of Hawking radiation being captured by the black hole, which means that the black  hole's ability to capture particles is high\cite{jyoti2023quasinormal,PhysRevD.14.3251}.

Therefore, in this Section we calculate the greybody factors for a Kerr--Taub--NUT black hole.

\subsection{GreyBody Factors}
We use the lower bound semi-analytic approach to calculate the GFs denoted by $\gamma_{\ell m} (\omega)$ , which has been used to establish limits in the particle production in different space-times\cite{al2020solution,boonserm2008bounding}. 

The expression to calculate the GFs is given by:

\begin{equation}
\label{gf}
\gamma_{\ell m} (\omega)  \geq \operatorname{sech}^2\left(\int_{-\infty}^{\infty} \vartheta \mathrm{d} r_*\right),
\end{equation}
where $\gamma_{\ell m} (\omega) $ is the greybody factor, and $\vartheta$ is the function:

\begin{equation}
\label{var} \vartheta=\dfrac{\sqrt{\left(h^{\prime}\right)^2+\left(\omega^2-V-h^2\right)^2}}{2 h}.
\end{equation}

Moreover, $h$ is some positive function, $h\left(r_*\right)>0$, satisfying the limits $h(-\infty)=h(+\infty)=\omega$, and $V$ is the effective potential taken from the Schrödinger--like wave equation  \eqref{wave1}.

Setting $h=\omega$ \cite{boonserm2008bounding}, Eq. \eqref{gf} turns into the following expression:

\begin{equation}
\label{gf1}
\gamma_{\ell m} (\omega)  \geq \operatorname{sech}^2\left(\frac{1}{2\omega}\int_{-\infty}^{\infty} V(r_{\ast}) \mathrm{d} r_*\right).
\end{equation}

Finally, getting out from the tortoise coordinate system,the final expression for the GFs is 

\begin{equation}
\label{gf2}
\gamma_{\ell m} (\omega)  \geq \operatorname{sech}^2\left(\frac{1}{2\omega}\int_{r_{H}}^{\infty} V(r) \mathrm{d} r\right).
\end{equation}

Once we have the effective potential of the KTN black hole, it is possible to calculate the greybody factors without solving the radial part of the Klein--Gordon equation \eqref{gordon1}, as we discussed in the previous section.
In order to apply Eq. \eqref{gf2}, let us return from the tortoise coordinate to the original coordinate. So that, the effective potential \eqref{potential} will be given in the following form:

\begin{eqnarray}
\label{potential1}
\nonumber
V_{\omega \ell m}(r)&=&\frac{2 ~a ~m~ \omega}{\Delta}-\frac{a^2 m^2- \ell (\ell+1)\Delta }{\Delta\left(a^2+l^2+r^2\right)}\\
&-&\frac{  r-  \left(2 r^2-2 M r\right)-\Delta }{\left(a^2+l^2+r^2\right)^2}-\frac{2 \Delta  r^2}{\left(a^2+l^2+r^2\right)^3}.
\end{eqnarray}
Now, using Eq. \eqref{gf2},~  we have the following expression
\begin{equation}
\label{gf4}
\gamma_{\ell,m} (\omega)  \geq \operatorname{sech}^2\left[\frac{1}{2\omega}\displaystyle\int_{r_{H}}^{\infty} V_{\omega \ell m}(r) \mathrm{d} r\right].
\end{equation}

Now solving the integral part of Eq. \eqref{gf4}, we arrive at

\begin{align}
\label{intpo}
\frac{1}{2\omega}\int_{r_{H}}^{\infty} V_{\omega \ell m}(r) \mathrm{d} r&= \dfrac{1}{8 \omega \left(M^2 \left(a^2+l^2\right)+l^4\right)}\\
\nonumber &\times\left(P_{\ell m}+Q_{m}(\omega)-R+S_{m}(\omega)+T_{m}\right),
\end{align}
where the functions $P_{\ell m}$,~ $Q_{m}(\omega)$,~ $R$,~ $S_{m}(\omega)$, ~ and~  $T_{m}$ were used in order to simplify the result of the integration and they are given by the following expressions:

\begin{eqnarray}
\nonumber
P_{\ell m} &=& \dfrac{1}{(a^2 + l^2)^{3/2}} \left[\pi - 2 \arctan\left(\frac{r_{h}}{\sqrt{a^2 + l^2}}\right)\right] \\
\nonumber
&\times& \left[a^4 l^2 m^2 + M^2 (a^2 + l^2) \left[2 a^2 (\ell^2 + \ell + 1) \right.\right.\\
\nonumber 
&+&\quad \left.\left.(2 \ell (\ell + 1) + 1) l^2\right]  + a^2 l^4 \left(2 \ell (\ell + 1) + m^2  +2\right)\right.\\
&+&\quad \left.\left(2 \ell \left(\ell + 1\right) + 1\right) l^6\right],\label{eq1}\\
\nonumber
Q_{m}(\omega)&=& \sqrt{\dfrac{1}{a^2-l^2-M^2}}a m \pi \left[4 \omega \left(M^2 \left(a^2+l^2\right)+l^4\right)\right.\\
&-& \left.a m \left(l^2+M^2\right)\right],\label{eq2}\\
\nonumber
\end{eqnarray}
\begin{eqnarray}
\nonumber
R&=&\dfrac{2 \left[M^2 \left(a^2+l^2\right)+l^4\right]}{\left(a^2+l^2\right) \left(a^2+l^2+r_{h}^2\right)^2} \left[\left(2 M+1\right) \left(a^2+l^2\right)^2\right.\\
& +&  \left.+r_{h}^2 \left(a^2+l^2\right)-3 l^2 r_{h} \left(a^2+l^2\right)-l^2 r_{h}^3\right],\label{eq3}\\
\nonumber
S_{m }(\omega)&=&\dfrac{2 a m  \arctan \left(\frac{M-r_{h}}{\sqrt{a^2-l^2-M^2}}\right)}{\sqrt{a^2-l^2-M^2}}\left[4 \omega \left(M^2 \left(a^2+l^2\right)+l^4\right)\right.\\
&-& \left. a m \left(l^2+M^2\right)\right],\label{eq4}\\
\nonumber
T_{m}&=&a^2 m^2 M \left[\log \left(a^2+l^2+r_{h}^2\right)\right.\\ 
& -& \left.\log \left(a^2-l^2-2 M r_{h}+r_{h}^2\right)\right].\label{eq5}
\end{eqnarray}

Therefore using the result  Eq. \eqref{intpo}, we can have the expression for the greybody factor for a KTN black hole, which is given as follows:

\begin{align}
\label{KTNg}
\gamma_{\ell m} (\omega)  \geq &\operatorname{sech}^2\left[\frac{1}{8 \omega \left(M^2 \left(a^2+l^2\right)+l^4\right)}\right.  \nonumber \\
&\times\left(P_{\ell m}+Q_{m}(\omega)-R+S_{m}(\omega)+T_{m}\right)\Bigg ].
\end{align}

After, analyzing Eq. \eqref{KTNg} it was possible to identify that the only allowed value for $m$ is zero; otherwise, the expression \eqref{KTNg} is indeterminate. Physically, when $m=0$ means that the  massless scalar field is dominated only by axisymmetric perturbations \cite{TeixeiradaCosta2020}.

Taking into account the previous fact for the value of $m$. The greybody factor \eqref{KTNg} will not depend on  the functions that depend on $m$. So that, Eq. \eqref{KTNg} can be expressed as follows:

\begin{equation}
\label{KTNg1}
\gamma_{\ell0} (\omega)  \geq \operatorname{sech}^2\left\{\dfrac{1}{8 \omega \left[M^2 \left(a^2+l^2\right)+l^4\right]}\left(P_{\ell 0}-R\right)\right\}.
\end{equation}
where $R$ is given by Eq. \eqref{eq3} and $P_{\ell 0}$ as:

\begin{eqnarray}
P_{\ell 0} &=& \dfrac{1}{(a^2 + l^2)^{3/2}} \left[\pi - 2 \arctan\left(\dfrac{r_{h}}{\sqrt{a^2 + l^2}}\right)\right] \nonumber \\
&\times& \left\{ M^2 (a^2 + l^2) \left[2 a^2 (\ell^2 + \ell + 1) + \left(2 \ell (\ell + 1) + 1\right) l^2\right] \right. \nonumber \\
&+&\left.  a^2 l^4 \left(2 \ell (\ell + 1)  + 2\right) + \left(2 \ell (\ell + 1) + 1\right) l^6\right\}.\label{on} 
\end{eqnarray}

\begin{figure*}[th]
\centering
\includegraphics*[width=1\textwidth]{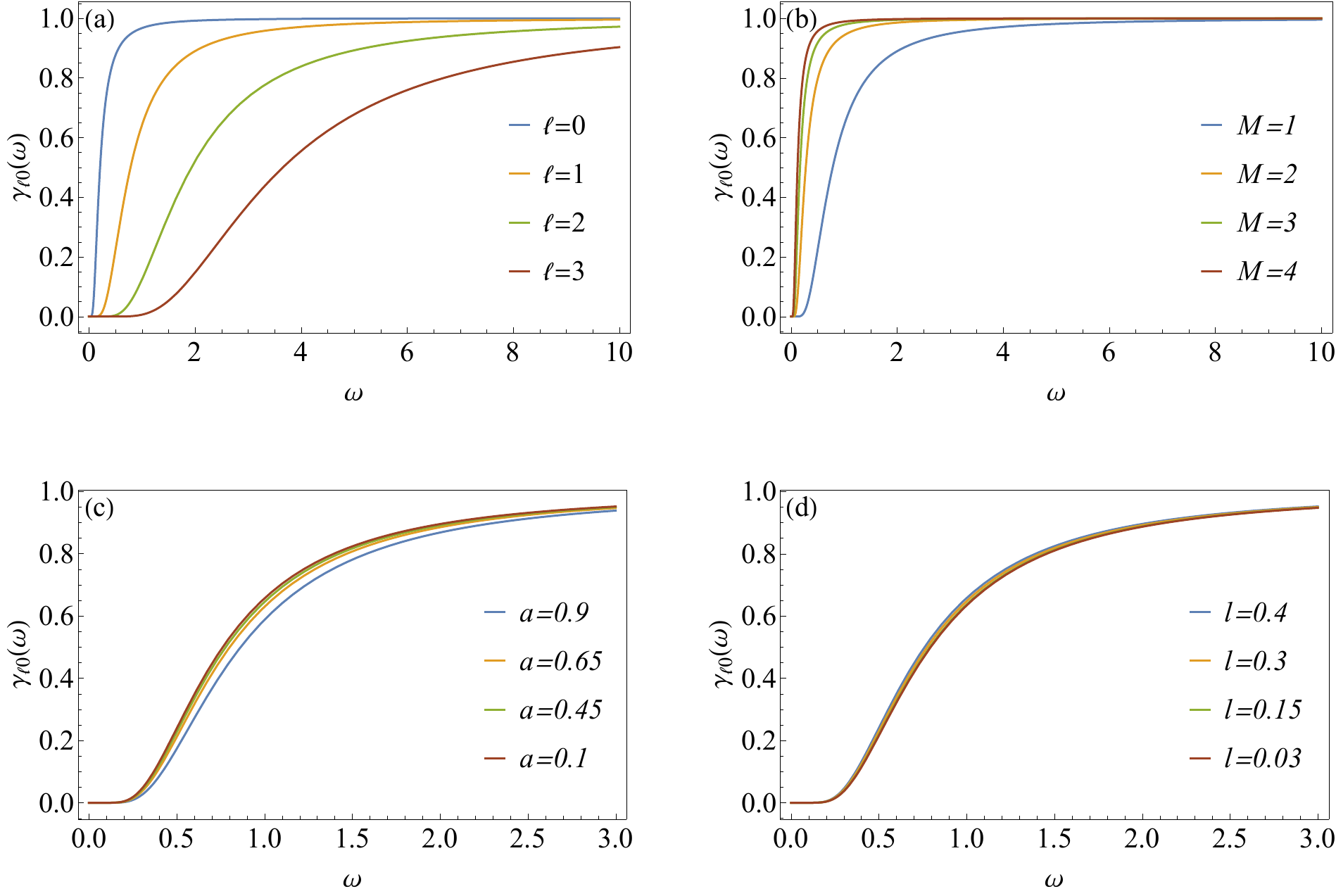}
\caption[Plots of the greybody factors $\gamma _{\ell 0}(\omega)$ \eqref{potential} for various situations. (a) $ M=1$, $a=0.5$, ~$l=0.3$, ~ (b) $\ell=1$,~ $a=0.5$, ~$l=0.3$,~  (c) $M=1$, ~$\ell=1$ ,~ $l=0.3$, ~ (d) $M=1$,~ $\ell=1$,~ $a=0.5$]{Plots of the greybody factors $\gamma _{\ell 0}(\omega)$ \eqref{potential} for various situations. (a) $ M=1$, $a=0.5$, ~$l=0.3$, ~ \quad (b) $\ell=1$,~ $a=0.5$, ~$l=0.3$, ~ (c) $M=1$, ~$\ell=1$ ,~ $l=0.3$, ~ (d) $M=1$,~ $\ell=1$,~ $a=0.5$}
\label{Fig8}
\end{figure*}

From figure \ref{Fig8} it is possible to analyze the behavior of the greybody factors for a KTN black hole. First of all, considering the change of the spheroidal harmonic quantum number $\ell$ Fig. \ref{Fig8}~(a), it is clear to see that increasing $\ell$ the GFs decrease, which means that less bosonic radiation will be captured by the black hole, instead of low $ \ell$ values. 

Looking at Fig. \ref{Fig8}~(b), it is possible to see that the GFs are affected by the black hole mass. It is clear that massive black holes have high GFs instead of low--mass black holes, where the GFs are small. It means that the black hole's ability to capture particles (i.e., Hawking radiation) is much more significant for high masses than low masses. As we discussed before, the less massive BHs have higher temperatures, and the GFs support it because the emission of particles is strong. In the other case, when the BHs are massive, their temperature is small because there is not much concentration of particles that escape from the black hole.

Analyzing the effect of the rotation parameter $a$ in the GFs Fig. \ref{Fig8}~(c), it is possible to identify that there is no effect of the rotation parameter in the GFs, which means that, despite the Kerr nature of the BH, the amount of radiation that comes away from the regions near the event horizon is approximately the same for all rotations. Practically, there is a small difference for low frequencies of the spin$-0$ particles. In the same way, considering the change in the NUT parameter Fig. \ref{Fig8}~(d), there is no effect in the GFs, which is an interesting behavior because we expected to have real variations in the GFs as a consequence of the NUT parameter. Moreover, the similarity between the GFs of the rotation and NUT parameters implies that the NUT parameter is related to a source of angular momentum, as suggested by Bonnor \cite{bonnor1969new}.

\section{Particle Emission Rates}
\label{PER}

Generally, as a consequence of the quantum fluctuations from the regions near the black hole horizon \cite{OVGUN2018138}, a pair of virtual particles is created, one of them with positive energy and the other with negative energy. The particle with negative energy falls into the black hole, and the other particle escapes to infinity as Hawking radiation \cite{schutz2003gravity}. This mechanism leads to black hole evaporation due to the negative contribution to the black hole mass\footnote[3]{We consider the equivalence between mass and energy.}, where the black hole parameters evolve over time.

As we discussed before, not all the positive energy particles or massless scalar perturbations can leave the black hole. Part of them are captured by the black hole due to the potential surrounding it. Therefore, the greybody factors will affect the energy emission rate, which gives the fraction of particles are captured by the black hole. 

When studying the Hawking radiation emitted by black holes, we will consider the assumption that the black hole is in a state of thermal equilibrium with its surroundings. Expressly, we assume that the temperature of the black hole remains constant between the emissions of two consecutive particles \cite{kanti2004black,sun2023quasinormal}. This assumption leads us to model the system using the canonical ensemble. Moreover, as we discussed before, GFs are valid for $m=0$, which implies that we do not need to consider terms related to $m$. Therefore, the Hawking radiation distribution is expressed as follows:

\begin{equation}
\label{black}
f=\dfrac{\gamma_{\ell 0}(\omega)}{e^{\frac{\omega}{T_{H}}}\pm1},
\end{equation}

\noindent where $\gamma_{\ell 0} (\omega)$ are the greybody factors, $T_{H}$ is the black hole's temperature, and the use of the $\pm$ sign depends on the nature of the particles that are going to be analyzed, for example, it is usual to use $+$ for fermions and $-$ for bosons.

In the same way, Hawking radiation is an essential quantum feature that carries information about the dynamics of black hole evaporation. Therefore, the energy emission rate of the black hole will depend on the parameters that describe the black hole; in that case, the dynamics of the mass, angular momentum, and NUT parameter, also known as the NUT charge, are essential to understand the Hawking radiation for the KTN black hole. To carry out our analysis, we will consider the model that Kokkotas \textit{et. al} \cite{kokkotas2011quasinormal} used to describe the emission rates for Kerr--Newman black holes in a magnetic field. Additionally, we need to sum for different modes of Eq. \eqref{black} using values of $\ell \leq 3$, because these values guarantee a real contribution to the radiation that is emitted by the black hole \cite{leite2019axis}. In the same way, we are going to take $m$ to evolve from 0 to the maximum value of $\ell$, so that, $0\leq\ell\leq 3$ and $0\leq m\leq 3$.

Moreover, it is important that to analyze the change in angular momentum, we need to analyze its behavior at infinity with respect to energy. At infinity, the wave is composed of many quanta, each with energy
$E=\omega $ and angular momentum in the $ \varphi$ direction $J=m$. Thus, the ratio of the total angular momentum to the total energy carried by the wave across a hypothetical sphere must be $\frac{m}{\omega}$ \cite{Brito2015}.

\subsection{Mass Emission Rate}

The energy emission rate is given as follows:

\begin{equation}
\label{energy}
-\dfrac{\mathrm{d} M}{\mathrm{d} t}=\sum_{\ell=0}^{3} \sum_{m=0}^3 \gamma _{\ell 0}(\omega) \dfrac{\omega}{\exp \left(\omega / T_H\right)-1} \dfrac{\mathrm{d} \omega}{2 \pi}.
\end{equation}

\begin{figure}[th]
\centering
\includegraphics*[width=0.48\textwidth]{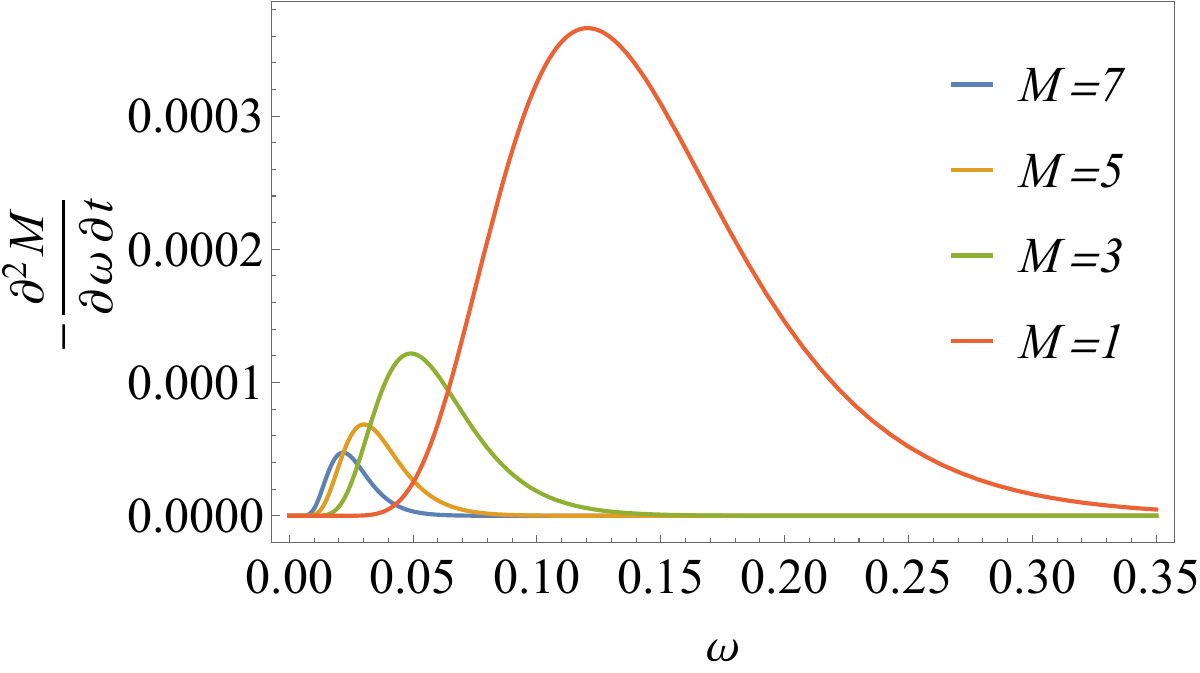}
\caption[Energy emission rate for KTN black hole for different masses $M$. We set the parameters $a=0.7$,~ $l=0.3$.]{Energy emission rate for KTN black hole for different masses $M$. We set the parameters $a=0.7$, ~$l=0.3$. }
\label{Fig9}
\end{figure}

From Figure \ref{Fig9}, it is clear that the emission rate changes depending on the black hole mass. First, the emission for massive black holes is very weak compared to low--mass black holes. If we see the greybody factors for changes in mass, Figure \ref{Fig8}~(b), the absorption probability is less than the massive black holes. It means that more particles are emitted during radiation when the holes are less massive. Moreover, the previous fact means that the lifetime of high--mass black holes is greater than the lifetime of low--mass black holes.

\subsection{Angular Momentum Emission Rate}

Taking into account the ratio of angular momentum with respect to the energy at infinity $m/ \omega$,~ the emission rate is defined as follows:

\begin{equation}
\label{momentum}
-\dfrac{\mathrm{d} J}{\mathrm{d} t}=\sum_{\ell=0}^{3} \sum_{m=0}^3 \int\ \gamma _{\ell 0}(\omega) \dfrac{m}{\exp \left(\omega / T_H\right)-1} \dfrac{\mathrm{d} \omega}{2 \pi}.
\end{equation}

\begin{figure}[th]
\centering
\includegraphics*[width=0.48\textwidth]{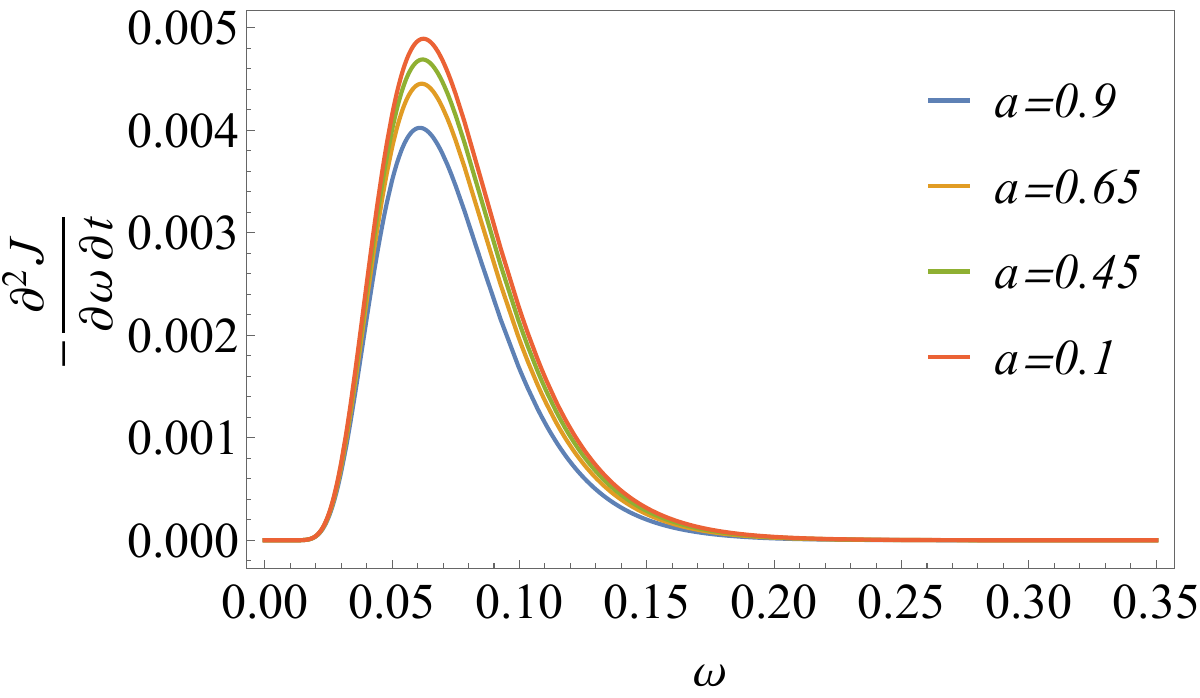}
\caption[Angular momentum emission rate for KTN black hole for different rotation parameters $a$. We set the parameters $M=2$,~ $l=0.3$.]{Angular momentum emission rate for KTN black hole for different rotation parameters $a$. We set the parameters $M=2$,~ $l=0.3$.}
\label{Fig10}
\end{figure}

From the data presented in Figure \ref{momentum}, noticeable variations in the angular momentum emission rate become apparent with rotation changes. Emission rates are higher for smaller rotations than their counterparts with higher rotations, where the rate tends to diminish.
At first glance, the emission should increase with a more significant rotation; however, as observed earlier, this is not the case. The primary reason is that particles find it challenging to escape as a black hole approaches its maximum rotation due to the significant dragging effect, causing it to rotate with the hole. Conversely, the dragging effect diminishes when the hole's rotation decreases, allowing particles to escape more easily from the event horizon. Thus, the level of rotation acts as a barrier, impeding radiation from reaching infinity.

\subsection{NUT Parameter Emission Rate}

Based on the model of Kokkotas \textit{et al.} \cite{kokkotas2011quasinormal}, we will assume a fundamental NUT charge $n$. This charge will be entirely within the theoretical framework. In general, let us assume that at infinity, a wave carries information about the NUT charge, which is quantized. Therefore, the total NUT charge $L$ will be defined as $L=n$. Analyzing the ratio of the total NUT charge to the total energy carried by the wave across a hypothetical sphere must be  $\frac{n}{\omega}$.

Therefore, the rate of the NUT charge will be defined as follows:

\begin{equation}
\label{charge}
-\dfrac{\mathrm{d} L}{\mathrm{d} t}=\sum_{\ell=0}^{3} \sum_{m=0}^3 \int\gamma _{\ell 0}(\omega) \dfrac{n}{\exp \left(\omega / T_H\right)-1} \dfrac{\mathrm{d} \omega}{2 \pi},
\end{equation}
where $L$ denotes the total NUT parameter of the system.

\begin{figure}[th]
\centering
\includegraphics*[width=0.48\textwidth]{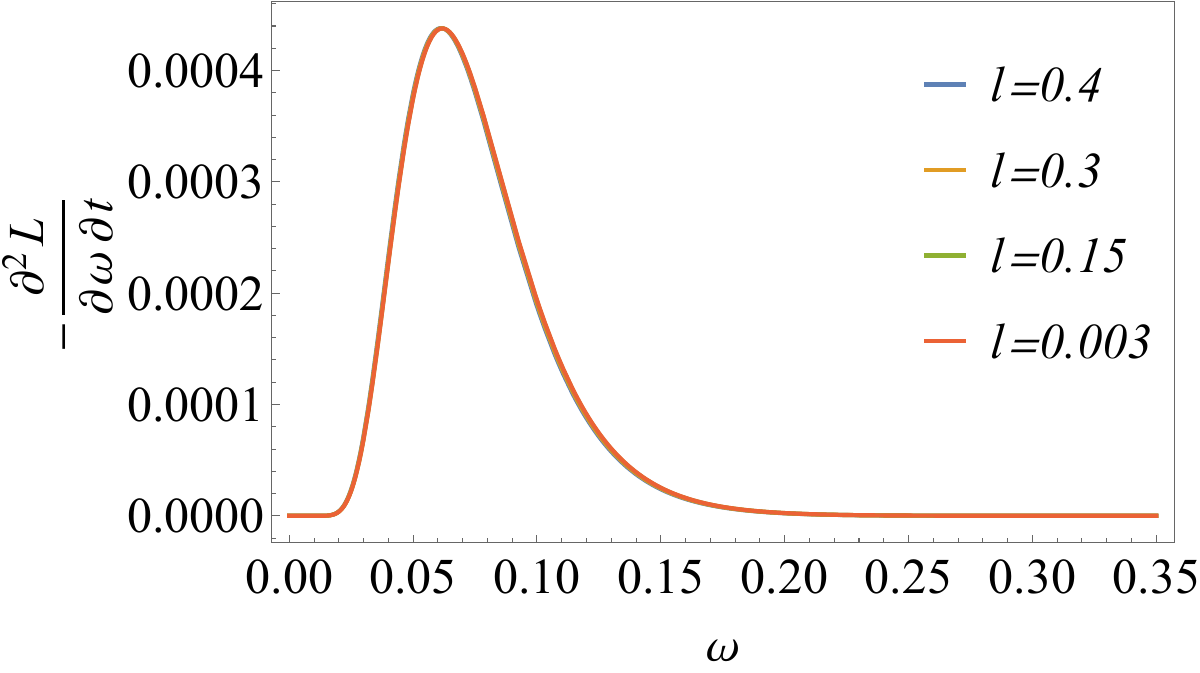}
\caption{NUT charge (NUT parameter) emission rate for KTN black hole for different NUT parameters $l$. We set the parameters $M=2$,~ $ a=0.7$, ~and ~$n=0.15$.}
\label{Fig11}
\end{figure}

In Figure \ref{Fig11}, it is possible to see that there is approximately no change in the emission due to the change of the NUT parameter. It means that particles will extract the NUT charge at a similar rate despite its different values. But, as we will see later, there is a small change in the emission rate for the NUT parameter.

\begin{figure}[th]
\centering
\includegraphics*[width=0.48\textwidth]{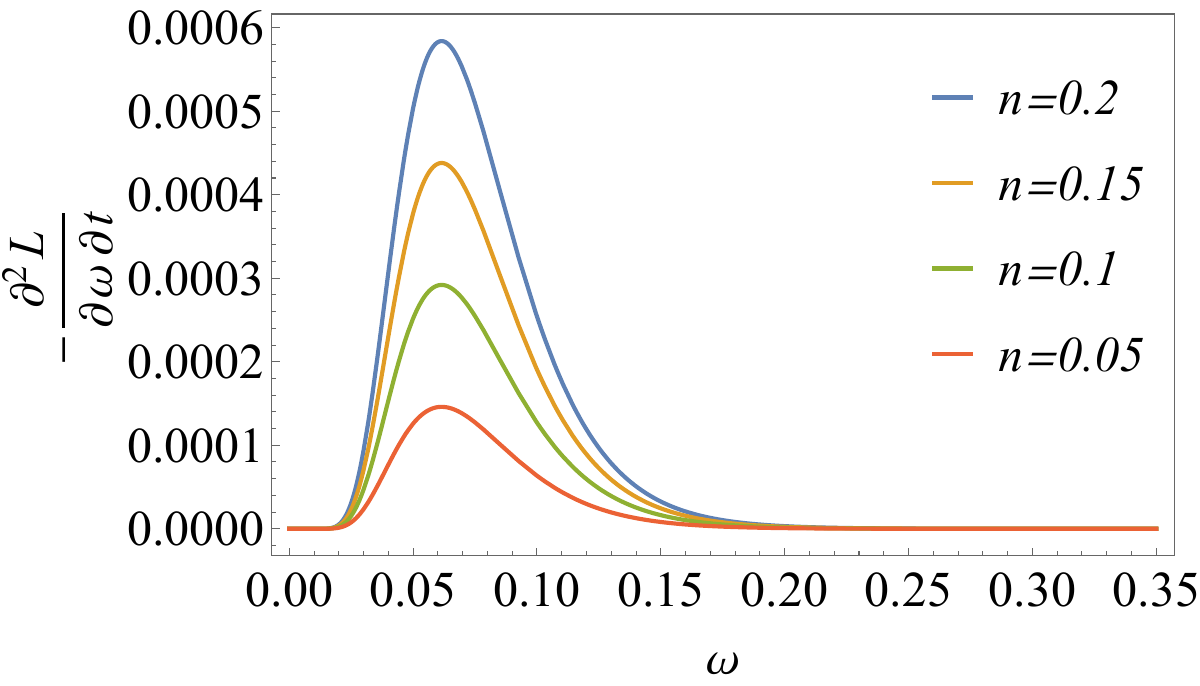}
\caption{NUT charge (NUT parameter) emission rate for KTN black hole for different NUT fundamental charges $n$. We set the parameters $M=2$,~ $ a=0.7$, ~and ~$l=0.3$.}
\label{Fignew}
\end{figure}

In Figure \ref{Fignew}, it is possible to see a change in the emission of the NUT parameter due to the fundamental NUT charge $n$. If the fundamental NUT charge has a high value, the emission is much faster than at low values for the fundamental NUT charge. Since the nature of the fundamental NUT charge is hypothetical, we think that studying the theoretical nature of $n$  would be interesting if we want to have a deeper understanding of the emission rate for the NUT parameter.

\subsection{Differential decays}

The differential decay practically quantifies the rate of the emission of a specific quantity\footnote[6]{The results were obtained integrating the data from the emission rates plots.}; in this case, the emission of mass, angular momentum, and the NUT parameter.

\begin{table*}[ht!]
\centering
\caption{Differential decays for the mass, angular momentum and NUT parameter. The fundamental NUT parameter is defined as $n=0.15$.}
\vspace{\baselineskip} 
\begin{tabular}{cccccc}
\toprule
Mass $M$& Rot. parameter $a$~ & NUT parameter $l$   & $-\dfrac{\mathrm{d}M}{\mathrm{d}t}=\alpha_{1}$& $-\dfrac{\mathrm{d}J}{\mathrm{d}t}=\alpha_{2}$ & $-\dfrac{\mathrm{d}l}{\mathrm{d}t}=\alpha_{3}$\\
\midrule
1 & 0.2 &0.10  &$8.9\times 10^{-5}$  & $9.3\times 10^{-4}$ & $9.3\times 10^{-5}$ \\
\hline
2& 0.3 & 0.20 & $1.6\times 10^{-5} $& $3.1\times 10^{-4} $& $3.1\times 10^{-5}$ \\
\hline
3& 0.4 & 0.30  & $6.2\times 10^{-6}$ & $1.8\times 10^{-4}$ & $1.8\times 10^{-5}$\\
\hline
4&  0.5& 0.35& $3.3\times 10^{-6}$ & $1.2\times 10^{-4}$  & $1.2\times 10^{-5}$\\
\hline
5 &  0.6& 0.40 &$2.1\times 10^{-6}$  & $9.6\times 10^{-5}$  & $9.6\times 10^{-6}$\\
\hline
6 & 0.7 &  0.50& $1.4\times 10^{-6}$ & $7.8\times 10^{-5}$  & $7.8\times 10^{-6}$\\
\bottomrule
\end{tabular}
\label{decays}
\end{table*}

From table \ref{decays}, it is possible to see that the differential decay for the mass is minor compared to the other decays, which means that the emission process for the mass is slow. In general, the evaporation for a KTN black hole is given by the following relation:

\begin{equation}
\label{relation}
-\dfrac{\mathrm{d}J}{\mathrm{d}t}>-\dfrac{\mathrm{d}L}{\mathrm{d}t}>-\dfrac{\mathrm{d}M}{\mathrm{d}t}.
\end{equation}

Based on relation \eqref{relation}, a Kerr--Taub--NUT black hole will first lose the angular momentum, then the NUT charge, and finally the mass. In other words, if we want to study a KTN black hole during different periods, it could be possible to neglect some parameters. For example, in our case, we can neglect the angular momentum and study a Taub--NUT black hole. From another point of view, after the angular momentum, we can avoid the NUT charge, which means that we are analyzing a Schwarzschild black hole. One example of this simplification was made by Page, where he neglected the charge of the black hole and considered only a Kerr black hole\cite{page1976particle}.

\section{Black Hole Parameters Time Evolution}\label{se4}

As we discussed in Section \ref{PER}, the parameters that describe the black hole change differently during the emission of Hawking radiation. Now, it is important to consider a model that describes the time evolution of the mass, angular  momentum, and NUT parameter.
In order to do that, it is important to consider the particle emission rates of Hawking radiation, which were discussed in Section \ref{PER} and the method used  by Page \cite{page2013time} and Nian \cite{nian2019kerr}, which can be generalized as follows
 \cite{page1976particle}:
 
\begin{equation}
\label{pagemodel}
\alpha=-M^3 \frac{\mathrm{d} \ln{X}}{\mathrm{d}t}.
\end{equation}

Eq. \eqref{pagemodel} is a model to calculate the particle emission rates \cite{Taylor1998}, where $X$ represents the parameter that needs to be considered for time evolution. Moreover, the luminosity or total power emitted is proportional to $M^{-2}$. As M decreases at this rate\cite{don13particle}, the black--hole lifetime will be proportional to $M^3$, which is the term that appears in Eq. \eqref{pagemodel}.

Depending on the parameters that need to be analyzed, model \eqref{pagemodel} gives an ordinary differential equation for each black hole parameter (mass, angular momentum, and the NUT parameter). Finally, to obtain the required time evolution of the black hole parameters (mass, angular momentum, and the NUT parameter), we need to use model \eqref{pagemodel} and form a system of differential equations that needs to be solved.

\bigskip
\noindent
For the mass;

\begin{equation}
\label{mass}
\dfrac{\mathrm{d}M(t)}{\mathrm{d}t}=-\dfrac{\alpha_{1}}{M(t)^2}.
\end{equation}

\bigskip
\noindent
For the angular momentum;

\begin{equation}
\label{moment}
\dfrac{\mathrm{d}J(t)}{\mathrm{d}t}=-\dfrac{\alpha_{2}a_{*}}{M(t)},
\end{equation}
where $a_{*}=\frac{J}{M^2}$.

\bigskip
\noindent
For the NUT parameter;

\begin{equation}
\label{nut}
\dfrac{\mathrm{d}L(t)}{\mathrm{d}t}=-\dfrac{\alpha_{3}l_{*}}{M(t)},
\end{equation}
where $l_{*}= \frac{l}{M}$ is the dimensionless NUT parameter.

The $\alpha_{1}, \alpha_{2}, \alpha_{3}$ come from the differential decays for every parameter of the black hole.

Combining equations \eqref{mass}, \eqref{moment}, and \eqref{nut} it is possible to have a system of differential equations. Additionally, in order to make the solution easy, let us consider the parameters $l_{*}$ and $a_{*}$ as independent variables \cite{page1976particle} and set them based on the initial conditions of the problem such that $a_{*}=\frac{J_{0}}{M_{0}^2}$ and $l_{*}=\frac{l_{0}}{M_{0}}$. 
Therefore, the solutions are:

\bigskip
For mass;

\begin{equation}
\label{solumass}
M(t)=(-3\alpha_{1}t+c_{1})^{1/3}.
\end{equation}

Now, taking into account that at $t=0$ we have an initial mass $M_{0}$ and that at the final step of the evaporation $M=0$, we have the following results;

\begin{equation}
\label{mo}
M(t)=(-3\alpha_{1}t+M_{0}^3)^{1/3},
\end{equation}
where the total time of decay is expressed as follows

\begin{equation}
\label{timed}
t_{decay}=\dfrac{M_{0}^3}{3\alpha_{1}},
\end{equation}
which describes the lifetime of the black hole. Additionally, it is clear to see that $t_{decay}$ is proportional to the initial mass of the black hole and inversely proportional to the rate of emission or differential decay. It means that, if the differential decay is small, the time will be very large, otherwise, the time will be small.

Therefore, the final expression for Eq. \eqref{solumass} is 

\begin{equation}
\label{finalm}
M(t)=M_{0}\left(1-\dfrac{t}{t_{decay}}\right)^{1/3}.
\end{equation}

As we did for the mass case, we applied the initial and final conditions for the NUT parameter and angular momentum \footnote{$J_0=J(0)$ is the initial angular momentum, and $L_0=L(0)$ is the initial NUT parameter.}. Therefore; 

For angular momentum;
\begin{equation}
\label{solumoment}
J(t)=J_{0}-\dfrac{\alpha_{2}a_{*}M_{0}^2}{2\alpha_{1}}\left[1-\left(1-\dfrac{t}{t_{decay}}\right)^{\sfrac{2}{3}}\right],
\end{equation}
where the time that angular momentum takes to decay is 

\begin{equation}
t_{momentum}=1-\left(\dfrac{\alpha_{2} a_{*} M_{0}^2-2 \alpha_{1} J_{0}}{\alpha_{2} a_{*} M_{0}^2}\right)^{\sfrac{3}{2}}~t_{decay}.
\end{equation}

\noindent
For the NUT parameter;

\begin{equation}
\label{solunut}
L(t)=L_{0}-\dfrac{\alpha_{3}l_{*}M_{0}^2}{2\alpha_{1}}\left[1-\left(1-\dfrac{t}{t_{decay}}\right)^{\sfrac{2}{3}}\right],
\end{equation}
where the time that the NUT charge takes to decay is 

\begin{equation}
t_{NUT}=1-\left(\dfrac{\alpha_{3} l_{*} M_{0}^2-2 \alpha_{1} L_{0}}{\alpha_{3} l_{*} M_{0}^2}\right)^{\sfrac{3}{2}}~t_{decay}.
\end{equation}

\begin{figure}[h]
\centering
\includegraphics*[width=0.48\textwidth]{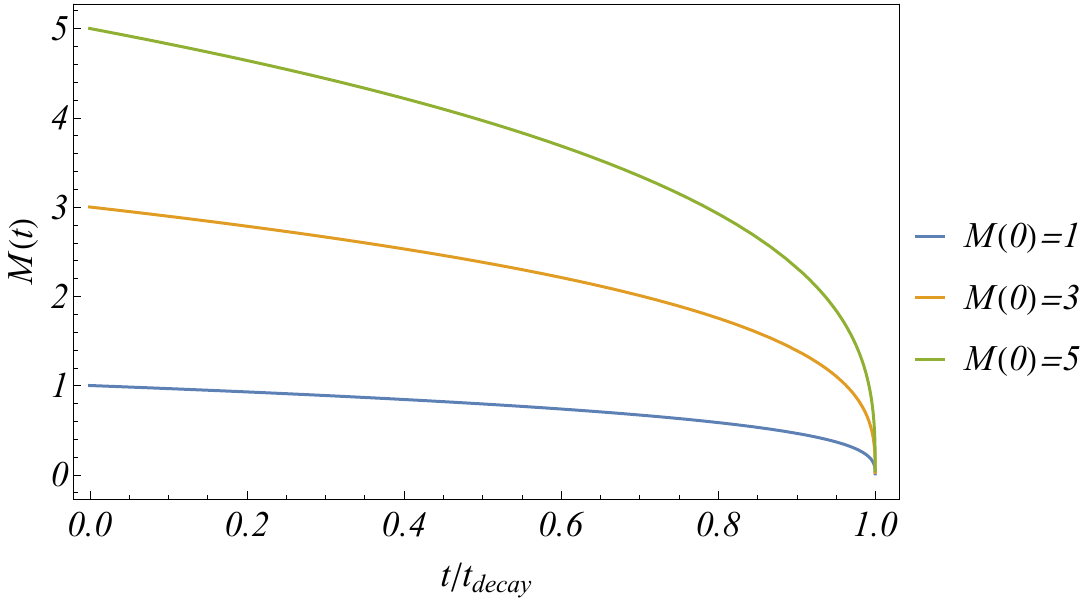}
\caption{Mass--time dependence profiles for different black hole masses.}
\label{Fig13}
\end{figure}

From figure \ref{Fig13}, it is possible to identify that massive BHs evolve more slowly than the less massive black holes. The previous fact makes sense because, as we discussed, the KTN black holes with small masses emit more radiation than the others, and as a consequence of this, the black hole will lose mass much faster than the massive ones.

Additionally, it is possible to get the same conclusion if we analyze the time of decay. Using the formula \eqref{timed}, we have the following results:

\begin{table}[h!]
\centering
\caption{Lifetime for different KTN black hole masses.}
\vspace{\baselineskip} 
\begin{tabular}{ccc}
\toprule
Mass $M$& $\alpha_{1}$& $t_{decay}/M^3$\\
\midrule
1   &$8.9\times 10^{-5}$  &$3.8\times 10^{3}$  \\
\hline
2 & $1.6\times 10^{-5} $ &$8.5\times 10^{4}$ \\
\hline
3   & $6.2\times 10^{-6}$  &$4.8\times 10^{5}$\\
\hline
4 & $3.3\times 10^{-6}$   & $1.6\times 10^{6}$\\
\bottomrule
\end{tabular}
\label{times}
\end{table}

Looking at the data from table \ref{times}, it is possible to confirm that massive black holes practically have a lifetime that is very long compared with the less massive black holes. 

\begin{figure}[h]
\centering
\includegraphics*[width=0.48\textwidth]{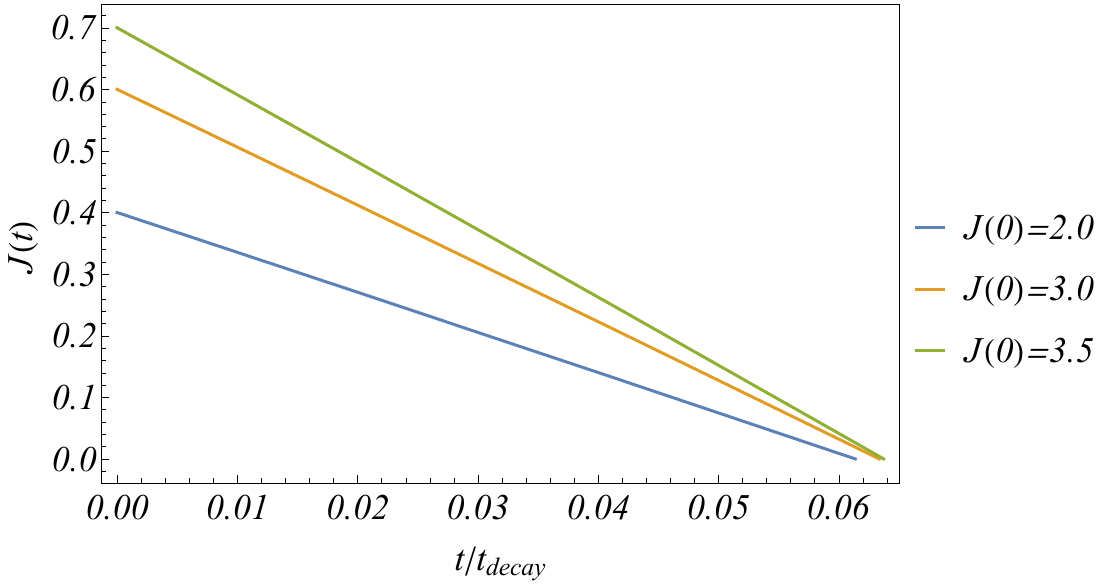}
\caption{Angular momentum--time dependence profiles for different black hole initial angular momentum. The physical parameters are $M=5$ and $l=0.6$}
\label{Fig14}
\end{figure}

Now, analyzing the evolution of the angular momentum \ref{Fig14}, it is possible to describe the following aspects: Firstly, the evolution of the angular momentum happens approximately linearly. In the same way, when we have small rotations, the angular momentum stays for less time; instead, when we increase the rotation, the angular momentum vanishes slowly. Physically, it means that when a black hole rotates slowly, the particles that escape from the event horizon regions extract rotational energy. As the rotation is slight, the black hole's angular momentum vanishes faster than when the rotations are high. As we discussed, it takes a lot of time to happen. In general, it means that when the rotation increases, it contributes slightly to the energy emission, and practically, depending on the amount of rotation, the angular momentum will eventually disappear.

\begin{figure}[h!]
\centering
\includegraphics*[width=0.48\textwidth]{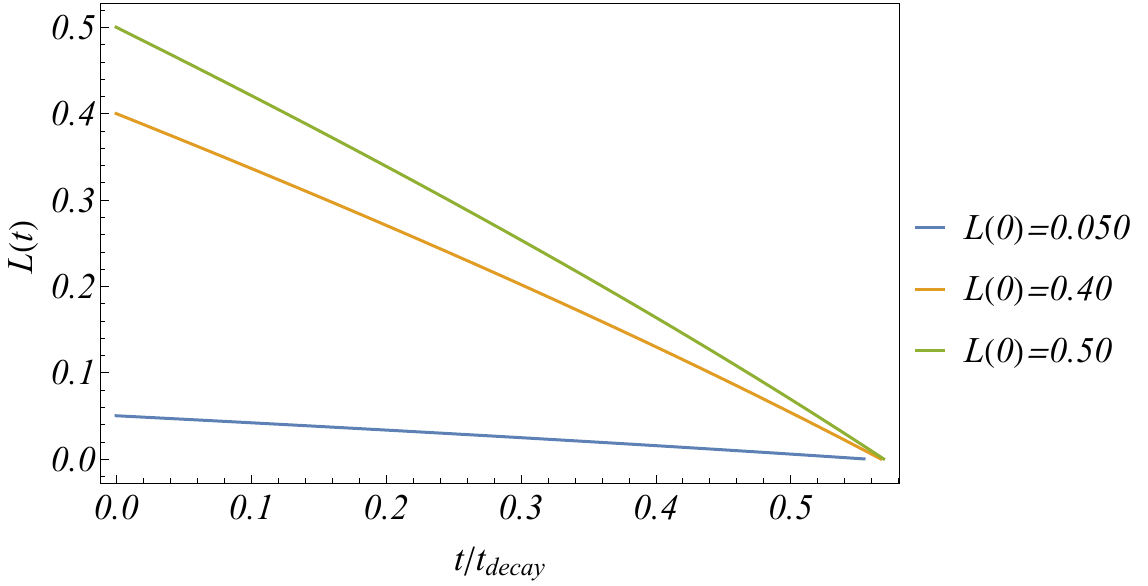}
\caption{NUT parameter time dependence profiles for different initial NUT values. The parameters are $M= 5$ and $a=0.6$.}
\label{Fig15}
\end{figure}

Note that analyzing the NUT parameter evolution (Fig. \ref{Fig15}) is similar to the angular momentum evolution. For small NUT values, the time it takes the NUT parameter to decay is less extended than in cases with high NUT parameters. In general, as in the case of angular momentum, the presence of NUT parameters affects its evolution. Physically, it implies that particles extract energy from the NUT parameter, which happens fast when it is small.

Now, comparing the evolution of angular momentum \eqref{Fig14} and NUT parameter \eqref{Fig15}, it is noticeable that the angular momentum has a shorter lifetime than the NUT parameter. This suggests that the percentage of the contribution of angular momentum to the energy emission is much more significant than the contribution of the NUT parameter, allowing a fast decay of the angular momentum.

\subsection{The Von Neumann Entropy}

Now, once we have the evolution of the parameters for a KTN black hole. It is possible to calculate the evolution of the Bekenstein--Hawking entropy.

Let us replace the equations \eqref{solumass}, \eqref{solumoment}, and \eqref{solunut} into expression \eqref{entropy2}, then we have

\begin{align}
\nonumber
S_{BH} (t) &= 2\pi\left.\Bigg[M(t)^2+L(t)^2\right.\\&+ \left.M(t) \sqrt{M(t)^2+L(t)^2-\dfrac{J(t)^2}{M(t)^2}}\right].\label{entropy3}
\end{align}

\begin{figure}[h!]
\centering
\includegraphics*[width=0.48\textwidth]{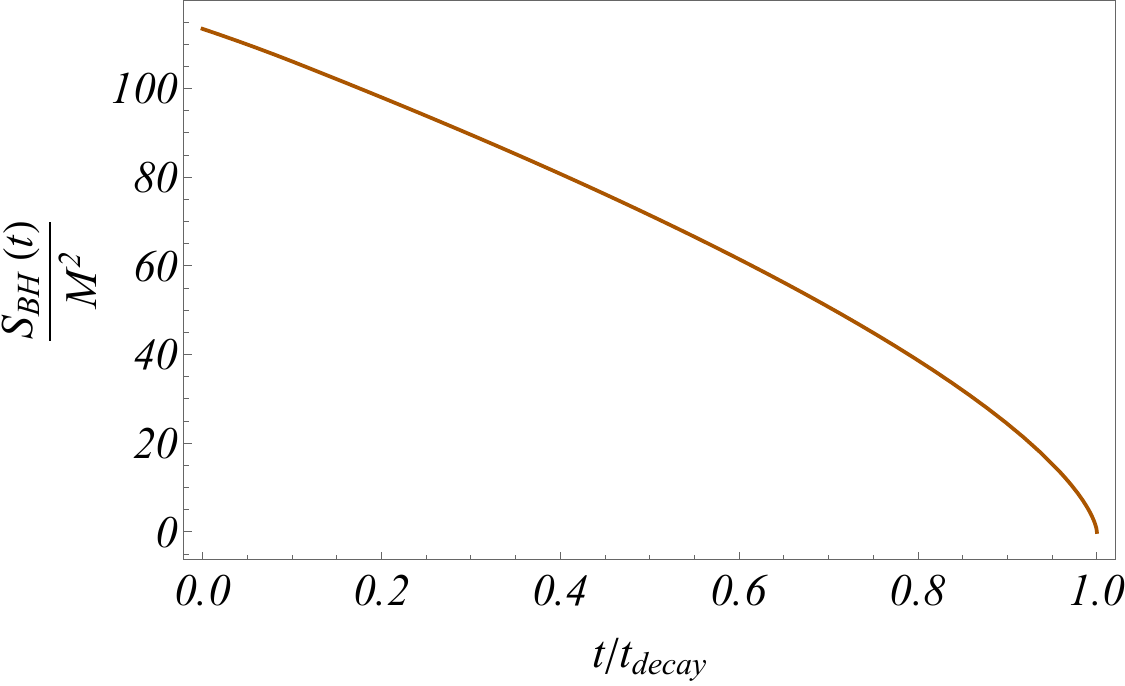}
\caption{Bekenstein--Hawking entropy evolution $S_{BH}(t)$ for a Kerr--Taub--NUT black hole. The physical parameters are:~$ M(0)=3$,~ $J(0)=1.2$,~ $ L(0)=0.3$.}
\label{Fig16}
\end{figure}

In figure \ref{Fig16}, we present the time dependence of the Bekenstein--Hawking entropy for a KTN black hole. As we discussed, this entropy states that the black hole
entropy is proportional to its event horizon (see subsection \ref{sarea}). Since the black hole is losing energy while emitting
radiation, it loses mass and shrinks until it is completely evaporated. Hence, this entropy begins at some value related to its initial conditions, and it decreases to zero \cite{mola2023black}.

Based on the the work of Nian \cite{nian2019kerr} and Page \cite{page2013time} the entropy of the radiation is given by the following expression:

\begin{equation}
\label{radentropy}
S_{rad}(t)=\beta \left[ S_{BH}(0)-S_{BH}(t)\right],
\end{equation}
where $S_{BH}$ is the Bekenstein--Hawking entropy, and $\beta$ is a parameter that characterizes the abundance of radiation. The parameter depends on the nature of the emitting particles. Since we are considering massless particles  \cite{page2005hawking}, $\beta$ is defined as the ratio between the change of radiation entropy, and black hole entropy during time. Thus,

\begin{equation}
\label{beta}
\beta =\dfrac{\frac{\mathrm{d}S_{rad}}{\mathrm{d}t}}{-\frac{\mathrm{d}S_{BH}}{\mathrm{d}t}}\approx 1.5003.
\end{equation}

Using \eqref{beta} as a reference, the radiation entropy is estimated to be 1.5003 times greater than the BH's reduction. Assuming that the black hole is solely producing photons, the previous number is derived from numerical calculations of Bekenstein--Hawking entropy and change of entropy of radiation \cite{page2005hawking,page2013time}. Generally speaking, $\beta$ computations are a current topic of research \cite{page2013time}. To observe the influence of $\beta$ on the radiation entropy and Von Neumann entropy, we will employ varying values of it.

\begin{figure}[h]
\centering
\includegraphics*[width=0.48\textwidth]{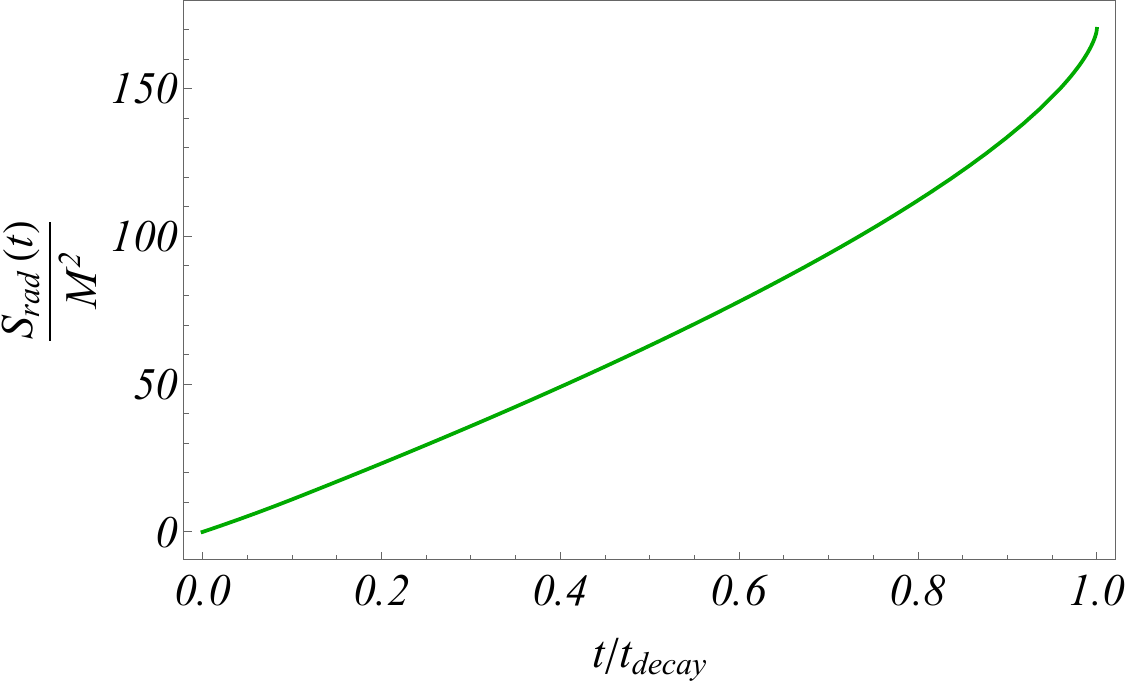}
\caption{Entropy of Hawking radiation evolution $S_{BH}(t)$ for a Kerr--Taub--NUT black hole. The physical parameters are:~$ M(0)=3$,~ $J(0)=1.2$, ~$L(0)=0.3$.}
\label{Fig17}
\end{figure}

In order to have the expression for the entropy of the Hawking radiation \eqref{radentropy}, it is only necessary to use the expression for the Bekenstein--Hawking entropy \eqref{bekenstein} and evaluate it at time $t=0$ and at time $t$. The calculations could be done manually or with some mathematical software. In figure \ref{Fig17}, we plotted the entropy of the Hawking radiation for a KTN black hole with mass $M=3$, angular momentum $J=1.2$, and NUT parameter $l=0.3$. The physical meaning is that just after the formation of the black hole, the entropy of the radiation is zero
since there is no radiation yet. Afterwards, the entropy increases as the Hawking radiation is
emitted until the black hole evaporates completely\cite{mola2023black, almheiri2020page,harlow2016jerusalem}. 

As we discussed in Section \ref{int}, the main purpose of this paper is to give a description about the Von Neumann entropy for a KTN black hole. In order to do this, we need to correlate the  black hole and its radiation since both must form a single quantum system. Using the work made by Page \cite{page2013time}, the entanglement entropy (i.e., Von Neumann entropy of the system) is given using the Heaviside step function because it correlates the radiation entropy and black hole entropy using the Page time as a threshold to put together the Bekenstein--Hawking (i.e., black hole entropy) and radiation entropies. Thus,

\begin{equation}
\label{ivon}
S_{VN}(t)= S_{rad}~\theta \left(t_{page}-t\right)+S_{BH}~\theta \left(t-t_{page}\right),
\end{equation}
where $t_{page}$ represents the time at which the thermodynamic entropy and the radiation entropy met. Therefore, in order to obtain it, we need to solve the following equation: $S_{rad}=S_{BH}$.

\begin{figure}[h!]
\centering
\includegraphics*[width=0.48\textwidth]{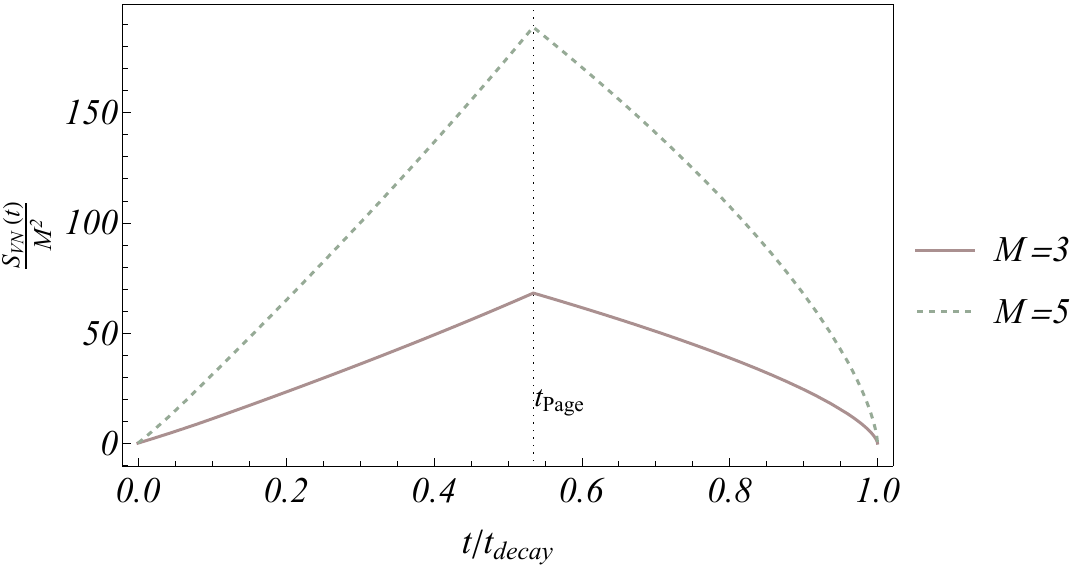}
\caption{Von Neumann entropy evolution for an initially pure KTN black hole. ~$M=3$,~ $J=1.2$, $l=0.3$ (thick line),~ $M=5$,~ $J=1.2$, ~$l=0.4$ (dashed line).}
\label{Fig20}
\end{figure}

The information extracted from figure \ref{Fig20} is described as follows: Firstly, if a KTN black hole together with its radiation is considered a pure quantum state, it means that there is a unitary mechanism that allows to conserve the information, as a pure state is evolving into a pure one. Secondly, as the Von Neumann entropy quantifies our lack of knowledge about a system, it is saying that at the end of the evaporation we have knowledge about the black hole information, which means information recovery. Additionally, considering the black hole parameters, it is clear that when the mass, angular momentum, and NUT parameter increase, it is more difficult to know the black hole information due to the higher values in comparison when the parameters (i.e., mass, angular momentum, and NUT parameter) are smaller.

Furthermore, despite the value of the black hole parameters, the time that it takes for the mass to be reduced to half is approximately the same \cite{karpasitis2021evaporating}. It is justified due to the Page time $t_{Page}=\frac{t}{t_{decay}}$ is similar for each situation; $t_{Page}=0.534653$  for a mass $M=3$, and $t_{Page}=0.534377$ for a mass $M=5$. 

\begin{figure}[h!]
\centering
\includegraphics[width=0.48\textwidth]{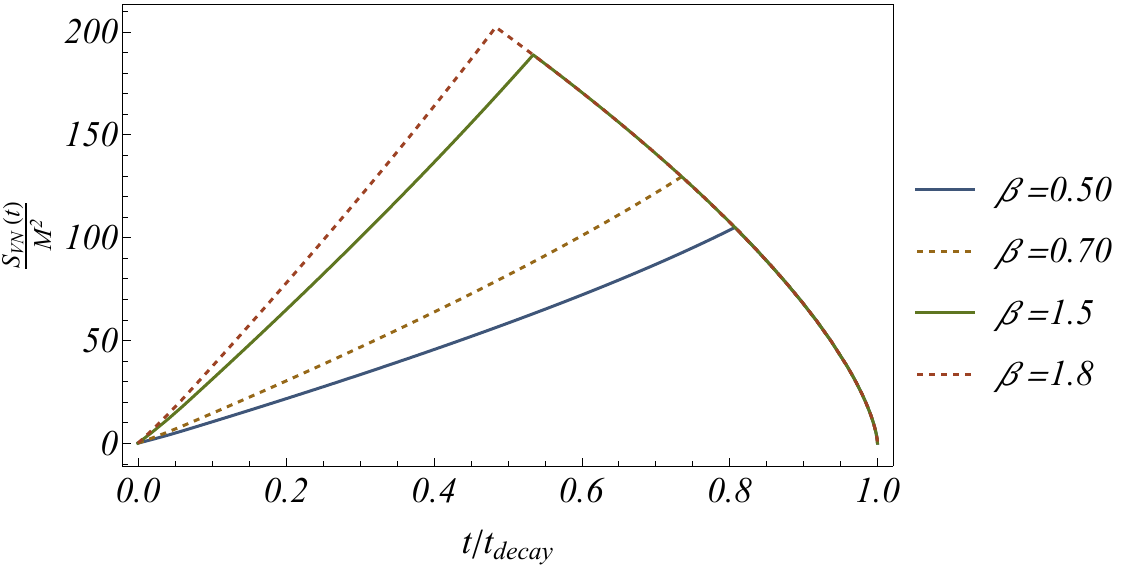}
\caption{Von Neumann entropy evolution for an initially pure KTN black hole for different $\beta$ values (see Eq. \eqref{beta}). We set $M=5$,~ $J=1.2$, ~$l=0.4$.}
\label{Fig23}
\end{figure}

As we discussed previously, we will assume arbitrary values for the  $\beta$ parameter in order to see its effects on the Page curve. From figure \ref{Fig23}, it is clear that when $\beta$ is small, it entails an increase in the Page time. It means that the black hole radiation takes more time in order to reduce the black hole mass to half. On the other hand, when the $\beta$ parameter increases, the Page time decreases. In general, it is possible to say that when the parameter is high, the process of evaporation is dominated by Hawking radiation, and the evaporation process takes less time compared with low values of $\beta$. On the other side, small values imply that practically there is a lack of radiation that allows a slow evaporation of the black hole. The Page time for the different values of $\beta $ is shown in table \ref{tpage}.

\begin{table}[h]
\centering
\caption{Page time for an initially pure KTN black hole for different $\beta$ values. $M=5$, $J=1.2$, $l=0.4$}
\vspace{\baselineskip} 
\begin{tabular}{c c}
\toprule
$\beta$ & Page time $\left[\dfrac{t}{t_{\text{decay}}}\right]$\\
\midrule
0.50  & $8.07\times 10^{-1}$\\
\hline
0.70  & $7.35\times 10^{-1}$\\
\hline
1.5  & $5.34\times 10^{-1}$\\  
\hline
1.8  & $4.83\times 10^{-1}$\\
\bottomrule
\end{tabular}
\label{tpage}
\end{table}

\section{Conclusions}
\label{se5}

Throughout this research, we made different calculations to get the evolution of the Von Neumann entropy for a Kerr--Taub--NUT black hole. In order to study Hawking radiation, we considered that the space-time surrounding the Kerr--Taub--NUT black hole is a massless scalar field. We studied the perturbations that the black hole generates in this field, and we analyzed the Klein--Gordon equation in a curved space \eqref{gordon1} by using the Teukolsky equation \eqref{radial}. We found that the effective potential that the black holes generate disappears at infinity due to its tendency to zero in the asymptotic limit. Moreover, an increase in the black hole rotation tends to make the potential broader and less intense.

In the same way, we studied the greybody factors, which represent the probability of a particle being absorbed by the black hole. From this, low--mass black holes represent less probability than massive black holes. In that sense, much more particles will be captured by the black hole in the second case. Thus, massive black holes have a higher probability of capturing particles. In the other case, small black holes have a lower probability of absorption, which guarantees a high number of particles that can reach infinity as Hawking radiation. Additionally, the rotation and NUT parameters have approximately the same absorption probability, which feeds the idea that the NUT parameter is related to a source of angular momentum. 

In black hole evaporation, we computed emission rates for the parameters characterizing the Kerr--Taub--NUT black hole. Our analysis indicated that the emission of angular momentum diminishes faster than the NUT charge and mass of the black hole, implying that the angular momentum diminishes faster than the other parameters. Eventually, during the final stages of evaporation, emissions align with those of a Schwarzschild black hole.

Finally, we used the model proposed by Page \eqref{pagemodel} to approximate the evolution of the KTN black hole parameters. After that, models \eqref{finalm},~\eqref{solumoment},~and ~\eqref{solunut}  were used to describe the evolution of the Bekenstein--Hawking and Hawking radiation entropy. We used the previous results to calculate the Von Neumann entropy that correlates the black hole and its radiation. We found that it follows the Page curve. It means that during the evaporation of a Kerr--Taub--NUT black hole, the information is preserved.
Additionally, we showed that regardless of the black hole parameters, the KTN black holes lose half of their initial mass at a similar time (i.e., Page time). Moreover, we analyzed the effect of the $\beta$ parameter, which quantifies the amount of Hawking radiation. The results suggested that high values of the $\beta$ parameter accelerate the evaporation process for a KTN black hole.

In general, we described the Von Neumann entropy for a Kerr--Taub--NUT black hole. This is important because the previous fact suggests that there is a unitary mechanism that guarantees that  information is not lost. Additionally, we obtain the result including terms only when $m=0$ (as a consequence of the GFs). Therefore, the previous fact represents a limitation of this work. Moreover, if the potential associated with the cosmic strings is included in the analysis, a change in results is expected. Thus, in future research, it will be essential to include the limitations mentioned above in order to refine the calculations and get a more detailed analysis of the Von Neumann entropy evolution.



\end{document}